\documentclass[aps,prb,twocolumn,superscriptaddress]{revtex4}
\usepackage{graphicx}
\usepackage{amssymb}
\usepackage{amsmath}
\usepackage{latexsym,bm,array,amsfonts,multirow}
\usepackage{epsfig}
\usepackage{color}
\usepackage{ulem}

\begin{document}

\title{Highly asymmetric superconducting dome and strange metallicity in La$_3$Ni$_2$O$_7$}
\author{Jiangfan Wang}
\email[]{jfwang@hznu.edu.cn}
\affiliation{School of Physics, Hangzhou Normal University, Hangzhou, Zhejiang 311121, China}
\author{Yi-feng Yang}
\email[]{yifeng@iphy.ac.cn}
\affiliation{Beijing National Laboratory for Condensed Matter Physics and Institute of
Physics, Chinese Academy of Sciences, Beijing 100190, China}
\affiliation{University of Chinese Academy of Sciences, Beijing 100049, China}
\affiliation{Songshan Lake Materials Laboratory, Dongguan, Guangdong 523808, China}
\date{\today}

\begin{abstract}
The recent discovery of high-temperature superconductivity in La$_{3}$Ni$_2$O$_7$ under high pressure has stimulated intensive investigations concerning its paring mechanism and a correct description of its effective low-energy physics. Notable experimental observations include the right-triangle-shaped superconducting dome with the maximum $T_c$ of about 80 K and the strange metal behavior in the normal state above $T_c$. Here we apply the Schwinger boson approach to the bilayer $t$-$V$-$J$ model, which allows us to treat well both the magnetic correlations and the superconducting instability. We obtain a global phase diagram with both metallic and superconducting ground states separated by a quantum phase transition, and predict a minimal interlayer superexchange $J$ and a minimal $d_{z^2}$-hole concentration necessary for the superconductivity, with a highly asymmetric right-triangle-like shape for the $T_c$ curve as well as a maximum $T_c$ comparable with experiments. The normal state is featured with a pseudogap in the $d_{x^2-y^2}$ spectra associated with preformed Cooper pairs and non-Fermi liquid strange metal behavior due to hybridization-induced spinon-holon-electron scattering. Our work clarifies the key difference of the two-component scenario of the superconductivity from other mechanisms, and provides a consistent understanding of the high-temperature superconductivity and strange metallic properties of La$_{3}$Ni$_2$O$_7$ under high pressure. 
\end{abstract}
\maketitle

\section{Introduction}

The newly discovered bilayer nickelate superconductor La$_{3}$Ni$_2$O$_7$ (LNO) with the maximum transition temperature  $T_c\approx 80$ K under high pressure has attracted great interest in clarifying its normal state and superconducting properties \cite{MWang2023Nature,JGCheng2023,HQYuan2024,XJZhou2024,HHWen2023,LShu2024,ZChen2024,XChen2024,ZXShen2024,JGCheng2024PRX,DLFeng2024,MWang2024,MWang2024b,HZLiu2024,LYang2024,LYang2024b,TXiang2024review,Yang2024arxiv}. Both experimental \cite{MWang2023Nature,XJZhou2024,HHWen2023} and theoretical studies \cite{DXYao2023,Dagotto2023,Werner2023,Leonov2023,Eremin2023,YYCao2024,Eremin2024} suggest that the low-energy physics is dominated by the $e_g$ orbitals of Ni ions, which form a bilayer structure with shared apical oxygen.  While the Ni-$d_{x^2-y^2}$ orbital is nearly quarter-filled and weakly correlated, the nearly half-filled Ni-$d_{z^2}$ orbital is more localized and strongly correlated \cite{XJZhou2024,Leonov2023,Eremin2023,YYCao2024}. The bilayer structure causes a strong interlayer superexchange interaction between the $d_{z^2}$ spins, which has been evidenced by the resonance inelastic X-ray scattering (RIXS) measurement \cite{DLFeng2024} and inelastic neutron scattering measurement \cite{MWang2024}. Upon applying pressure, the system undergoes a structure transition with an abrupt change of the interlayer Ni-O-Ni bond angle from 168$^\circ$ to 180$^\circ$ and a shortened $c$-axis lattice constant \cite{MWang2023Nature,HZLiu2024,MWang2024b}, which may further enhance the interlayer superexchange interaction and drive the pressurized superconductivity \cite{YFYang2023,QQin2023,GMZhang2023,WWu2023,DXYao2023tJ,TZhou2023,CJWu2023,JPHu2023,FYang2023,QHWang2023,GSu2023,CJWu2024,YHZhang2023,Bohrdt2023,GSu2023b,YZYou2023,ZYWeng2023,ZYLu2024,FWang2024,Kuroki2024,Kuroki2024b,WLi2024}. Although cuprate-like intralayer pairing mechanisms have also been proposed \cite{TXiang2023,KJiang2024,WKu2024}, the observed opposite trend of the maximum $T_c$ that is reduced from about 80 K in the bilayer LNO to about 30 K in the trilayer La$_4$Ni$_3$O$_{10}$ seems to support the interlayer pairing scenario in these multilayer nickelate superconductors \cite{JZhao2024,QQin2024,Yang2024PRB}.   

Among all interlayer pairing scenarios, there are two major proposals distinguished by the different roles of $d_{x^2-y^2}$ and $d_{z^2}$ orbitals. One is the two-component theory \cite{YFYang2023,QQin2023}, which suggests that the local interlayer spin singlet valence bonds between $d_{z^2}$ electrons provide the pairing, while the metallic $d_{x^2-y^2}$ electrons help to establish the phase coherence through their hybridization. The other emphasizes the strong Hund's rule coupling, which passes the interlayer superexchange of $d_{z^2}$ spins to the $d_{x^2-y^2}$ electrons, leading to an effective one-band ($d_{x^2-y^2}$) description of LNO \cite{GSu2023,YHZhang2023,Bohrdt2023,CJWu2024}. Numerical studies suggest that the hybridization and Hund's coupling may indeed induce distinct superconducting phases \cite{GSu2023b}. It is therefore important to clarify which scenario may be relevant for LNO. One important clue from the density-functional-theory (DFT) calculations is that the $\gamma$ band of the $d_{z^2}$ bonding orbital acquires a hole pocket at high pressure concurrently with the emergence of superconductivity \cite{MWang2023Nature,DXYao2023}, indicating the key role of $d_{z^2}$ holes for the superconducting order. In contrast, the Hund's coupling scenario predicts that the $d_{z^2}$ holes are detrimental to the superconductivity \cite{GSu2023b}. Other clues include the non-Fermi liquid (NFL) strange metal behaviors  \cite{HQYuan2024} and the recently observed pseudogap-like feature in the normal state \cite{LYang2024}. A proper low-energy effective theory should be able to address all these important properties of LNO.

In this work, we study the bilayer $t$-$V$-$J$ model proposed in the two-component theory, where $V$ represents the effective hybridization and $J$ represents the interlayer superexchange between $d_{z^2}$ spins \cite{YFYang2023}.  We develop an efficient method to calculate the Cooper instability within the Schwinger boson (slave fermion) representation of this model. The resulting phase diagram reveals a quantum phase transition between a hybridized metal with highly delocalized $d_{z^2}$ electrons and a superconducting phase. By tuning the $d_{z^2}$ occupation, we find a minimal $d_{z^2}$ hole concentration for the formation of superconducting order and a maximum $T_c$ consistent with experiment. In the normal state above $T_c$, we observe a pseudogap feature in the density of states (DOS) of the $d_{x^2-y^2}$ electrons, which occurs when the $d_{z^2}$ spins form interlayer singlet bonds for preformed Cooper pairs. Within a large temperature range, the imaginary part of the $d_{x^2-y^2}$ self-energy shows a $T^\alpha$ dependence with $1\leq \alpha < 2$, indicating NFL  behaviors in the normal state. We attribute these unusual features to the hybridization-induced spinon-holon-electron scattering. Our work thus provides a consistent picture of the major experimental observations of LNO under pressure and proposes potential experiments to distinguish the two-component theory of the superconductivity from other scenarios.

\section{Model and Methods}

\textit{Model.---}We use the effective bilayer $t$-$V$-$J$ model to describe the low-energy physics of LNO \cite{YFYang2023,QQin2023}:
\begin{eqnarray}
	H&=&-\sum_{lijs}(t_{ij}+\mu\delta_{ij})c_{lis}^{\dagger}c_{ljs}-\sum_{lijs}V_{ij}\left(c_{lis}^{\dagger}d_{ljs}+h.c\right) \label{eq:H} \nonumber\\
	&&+J\sum_{i}\bm{S}_{1i}\cdot\bm{S}_{2i}, 
\end{eqnarray}
where $c_{lis}^\dagger$ ($d_{lis}^\dagger$) creates a $d_{x^2-y^2}$ ($d_{z^2}$) electron on site $i$ of the $l$-th ($l=1,2$) layer, $s=\pm 1$ denotes the up and down spins, $\bm{S}_{li}=\frac12\sum_{ss'}d_{lis}^{\dagger}\bm{\sigma}_{ss'}d_{lis'}$ is the $d_{z^2}$ spin density operator, $V_{i,i+\hat{x}}=-V_{i,i+\hat{y}}=V/2$ is the in-plane hybridization between $d_{z^2}$ and $d_{x^2-y^2}$ orbitals on nearest-neighbor sites, and $J$ is the interlayer superexchange between $d_{z^2}$ spins. For general purpose, we only consider the nearest-neighbor hopping of $d_{x^2-y^2}$ electrons, set $t_{\langle ij\rangle}=t=1$ as the energy unit, and take the chemical potential $\mu=-1.44$ to give a quarter filling at $V=0$. As a minimal effective model, the above Hamiltonian ignores other small parameters such as the intralayer superexchange interaction, which is shown by RIXS and neutron scattering experiments to be much smaller than the interlayer one \cite{DLFeng2024,MWang2024}. The Hund's rule coupling is not explicitly included, but a part of its effect has been considered in the renormalized $V$ \cite{YYCao2024,YFYang2023}.

\textit{Schwinger boson method.---}We use the Schwinger boson (slave fermion) representation of the strongly correlated $d_{z^2}$ electrons \cite{Yoshioka1989,Long2022,Long2023,Wang2020,Wang2021,Wang2022a,Wang2022b}, $d_{lis}=\chi_{li}^\dagger b_{lis}$, where the slave fermion operator $\chi_{li}^\dagger$ creates a $d_{z^2}$ hole (holon), the Schwinger boson $b_{lis}^\dagger$ creates a $d_{z^2}$ spin (spinon), and the double occupancy has been projected out due to the large onsite Coulomb repulsion. The slave particles satisfy the constraint $Q_{li}\equiv\chi_{li}^\dagger \chi_{li}+\sum_s b_{lis}^\dagger b_{lis}=1$, which can be implemented via the Lagrange multiplier term $H_{\lambda}=\sum_{li}\lambda_{li}\left(Q_{li}-1\right)$. Since the $d_{z^2}$ occupation number is $\sum_s\langle d_{lis}^\dagger d_{lis}\rangle=\sum_s\langle b_{lis}^\dagger b_{lis}\rangle\equiv n_b$, the holon density  $n_\chi\equiv\langle \chi_{li}^\dagger \chi_{li}\rangle=1-n_b$ also counts the self-doped $d_{z^2}$ holes. 

In this representation, the $d_{z^2}$ spin operator is written as $\boldsymbol{S}_{li}=\frac{1}{2}\sum_{ss'}b_{lis}^\dagger \boldsymbol{\sigma}_{ss'}b_{lis'}$, and the superexchange term can be decomposed as,
\begin{eqnarray}
	J\boldsymbol{S}_{1i}\cdot \boldsymbol{S}_{2i}&=&-\frac{J}{2}\sum_{ss'}\left(sb_{1is}^\dagger b_{2i,-s}^\dagger\right)\left( s' b_{2i,-s'}b_{1is'}\right)+C \notag \\
	&\rightarrow&\Delta_i\sum_s s b_{1is}^\dagger b_{2i,-s}^\dagger+c.c.+\frac{2|\Delta_i|^2}{J}, \label{eq:HJ}
\end{eqnarray}
where $C$ is an irrelevant constant and $\Delta_i$ is the interlayer spin-singlet valence bond field. The hybridization term becomes a three-particle interacting vertex:
\begin{eqnarray}
	-\sum_{lijs}V_{ij}c_{lis}^\dagger d_{ljs}=-\frac{V}{\sqrt{\mathcal{N}_s}}\sum_{l{\bf kp}s}c_{l\mathbf{k}s}^\dagger \chi_{l\mathbf{p}}^\dagger b_{l,\mathbf{p}+\mathbf{k},s}\xi_\mathbf{k}, \label{eq:Hv}
\end{eqnarray}  
where $\mathcal{N}_s$ is the lattice size, and $\xi_\mathbf{k}=\cos k_x-\cos k_y$ comes from the sign structure of the nearest-neighbor hybridization. 

\begin{figure}[t]
	\begin{center}
		\includegraphics[width=8.5cm]{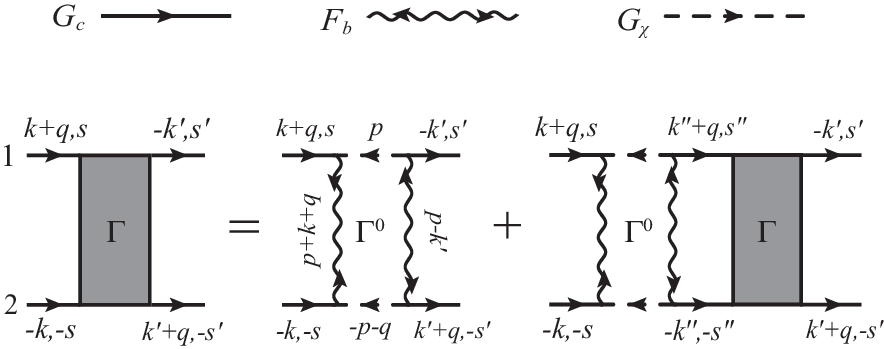}
	\end{center}
	\caption{A diagrammatic representation of the Bethe-Salpeter equation for the interlayer pairing vertex of $d_{x^2-y^2}$ electrons.}
	\label{fig:FD}
\end{figure}

To proceed, we first take the mean-field approximation assuming $\Delta_{i}=\Delta$ and $\lambda_{li}=\lambda$. The resulting action is 
\begin{eqnarray}
	S&=&-\sum_{lks}\bar{c}_{lks}(i\omega_n-\epsilon_{\bf k})c_{lks}-\sum_{lks}\bar{b}_{lks} (i\nu_n-\lambda)b_{lks} \notag \\
	&&+\Delta\sum_{ks}s\bar{b}_{1ks}\bar{b}_{2,-k,-s}+c.c.-\sum_{lk}\bar{\chi}_{lk} (i\omega_n-\lambda)\chi_{lk}\notag\\
	&&-\frac{V}{ \sqrt{\beta\mathcal{N}_s}}\sum_{lpks}\bar{c}_{lks} \bar{\chi}_{lp} b_{l,p+k,s}\xi_{\bf k}+c.c. \notag \\
	&&+2\beta\mathcal{N}_s\left(\frac{|\Delta|^2}{J}-\lambda\right), \label{eq:S}
\end{eqnarray}
where $\beta =1/T$ is the inverse temperature, $\epsilon_{\bf k}$ is the $d_{x^2-y^2}$-electron dispersion, and we have concisely written $k\equiv(\mathbf{k},i\omega_n)$ for fermions and $k\equiv(\mathbf{k},i\nu_n)$ for bosons. The above action is similar to that of the Kondo-Heisenberg model under the Schwinger boson representation \cite{Wang2020,Wang2021,Wang2022a,Wang2022b}, but with different holon dynamics.

The Green's functions of $d_{x^2-y^2}$ electrons, holons and spinons can be derived as:
\begin{eqnarray}
	G_c(\mathbf{k},i\omega_n)&=&\frac{1}{i\omega_n-\epsilon_\mathbf{k}-\Sigma_c(\mathbf{k},i\omega_n)}, \notag \\
	G_\chi(i\omega_n)&=&\frac{1}{i\omega_n-\lambda-\Sigma_\chi(i\omega_n)},\notag \\
	G_b(i\nu_n)&=&\frac{\gamma_b(-i\nu_n)}{\gamma_b(i\nu_n)\gamma_b(-i\nu_n)-|\Delta|^2}, \notag \\
	F_b(i\nu_n)&=&\frac{\Delta}{\gamma_b(i\nu_n)\gamma_b(-i\nu_n)-|\Delta|^2}, \label{eq:GG}
\end{eqnarray}
where 
$F_b(i\nu_n)=-s\langle b_{1ks}b_{2,-k,-s}\rangle$ is the anomalous Green's function of spinons, and $\gamma_b(i\nu_n)=i\nu_n-\lambda-\Sigma_b(i\nu_n)$. Note that holons and spinons are both local due to the lack of bare dispersions, while the $d_{x^2-y^2}$-electron Green's function is momentum dependent. The one-loop approximation gives the following self-energies:
\begin{eqnarray}
	\Sigma_c(\mathbf{k},i\omega_n)&=&\xi_\mathbf{k}^2\frac{V^2}{\beta} \sum_{m}G_\chi(i\omega_m)G_b(i\omega_m+i\omega_n), \notag \\
	\Sigma_{\chi}(i\omega_n)&=&\frac{2V^2}{\beta}\sum_m \bar{G}_c(i\omega_m)G_b(i\omega_m+i\omega_n), \label{eq:SE} \\
	\Sigma_b(i\nu_n)&=&-\frac{V^2}{\beta}\sum_{m}\bar{G}_c(i\omega_m)G_\chi(i\nu_n-i\omega_m), \notag
\end{eqnarray}
where we have defined $\bar{G}_c(i\omega_n)=\mathcal{N}_s^{-1}\sum_\mathbf{k}G_c(\mathbf{k},i\omega_n)\xi_\mathbf{k}^2$. One can derive Eq. (\ref{eq:SE}) rigorously by first using the large-$N$ expansion ($N$ being the number of spin components) and then setting $N=2$ \cite{Wang2020,Wang2021,Wang2022a,Wang2022b}. The mean-field equations for $\Delta$ and $\lambda$ can be obtained by minimizing the free energy with respect to $\Delta$ and $\lambda$, which gives
\begin{eqnarray}
	1&=&-\frac{2}{\beta}\sum_nG_b(i\nu_n)+\frac{1}{\beta}\sum_nG_\chi(i\omega_n),\notag \\
	\frac{\Delta}{J}&=&\frac{1}{\beta}\sum_nF_b(i\nu_n). \label{eq:constr}
\end{eqnarray} 
The self-consistent equations (\ref{eq:GG})-(\ref{eq:constr}) can be solved very efficiently in the imaginary-time space using the fast Fourier transform algorithm.

\textit{Cooper instability.---}We study the interlayer pairing instability of $d_{x^2-y^2}$ electrons by calculating the four-point correlation function $C_{ss'}(k,k',q)=\langle \bar{c}_{1,k+q,s} \bar{c}_{2,-k,-s} c_{2,k'+q,-s'}c_{1,-k',s'}\rangle$, where we have denoted $k\equiv (\mathbf{k},i\omega_n)$, $k'\equiv (\mathbf{k}',i\omega_{n'})$, and $q\equiv (\mathbf{q},i\nu_l)$. Using perturbative method, one can expand the correlation function into a series of  ladder diagrams and obtain the Bethe-Salpeter equation for the pairing vertex \cite{Simons2010} as shown in Fig. \ref{fig:FD}:
\begin{align}
	\Gamma_{ss'}&(k,k',q)= \Gamma_{ss'}^0(k,k',q)+\frac{1}{\beta\mathcal{N}_s}\sum_{k''s''}\Gamma_{s''s'}(k'',k',q) \notag \\
	&\times G_c(k''+q)G_c(-k'')\Gamma_{ss''}^0(k,-k''-q,q), \label{eq:Gamma}
\end{align}
where
\begin{align}
	\Gamma_{ss'}^0&(k,k',q)=ss'\frac{V^4 }{\beta\mathcal{N}_s}\sum_{p}F_b(p+k+q)^*F_b(p-k') \notag \\
	&\times G_\chi(p)G_\chi(-p-q)\xi_\mathbf{k+q}\xi_{-\mathbf{k}}\xi_{\mathbf{k}'+\mathbf{q}}\xi_{-\mathbf{k}'}  \label{eq:Gamma0}  
\end{align}
is the ``bare'' pairing vertex. The above equations suggest that the formation of superconducting order requires the interlayer spinon valence bond to provide the pairing strength, and the $d_{z^2}$ charge (holon) fluctuations and $d_{x^2-y^2}$-electron propagation to establish the superconductivity, which reflects precisely the two-component picture of the superconductivity in LNO \cite{YFYang2023}.

Equations (\ref{eq:Gamma}) and (\ref{eq:Gamma0}) can be further simplified by considering only the static uniform ($\nu_l=0$, $\mathbf{q}=\mathbf{0}$) solution, which leads to  $\Gamma_{ss'}(\mathbf{k},i\omega_n;\mathbf{k}',i\omega_{n'};\mathbf{0},0)=ss'\tilde{\Gamma}(i\omega_n,i\omega_{n'})\xi_\mathbf{k}^2\xi_{\mathbf{k}'}^2$. The form factor $\xi_\mathbf{k}^2\xi_{\mathbf{k}'}^2$ suggests that the $d_{x^2-y^2}$ paring in the momentum space contains nodes along the diagonal direction of the two-dimensional Brillouin zone, in agreement with previous studies \cite{YFYang2023,QQin2023}. The development of Cooper instability corresponds to the divergence of $\tilde{\Gamma}(i\omega_n,i\omega_{n'})$. Our calculations show that $\text{Re}\tilde{\Gamma}(i\omega_n,i\omega_{n'})$ is peaked at the smallest frequency, $|\omega_n|=|\omega_{n'}|=\pi T$, and decays very fast as the frequency increases, while $\text{Im}\tilde{\Gamma}(i\omega_n,i\omega_{n'})$ is almost zero for all frequencies. Therefore, the superconducting transition temperature is determined through the divergence of $\text{Re}\tilde{\Gamma}(i\pi T,i\pi T)$. 

\begin{figure}[t]
	\begin{center}
		\includegraphics[width=8.8cm]{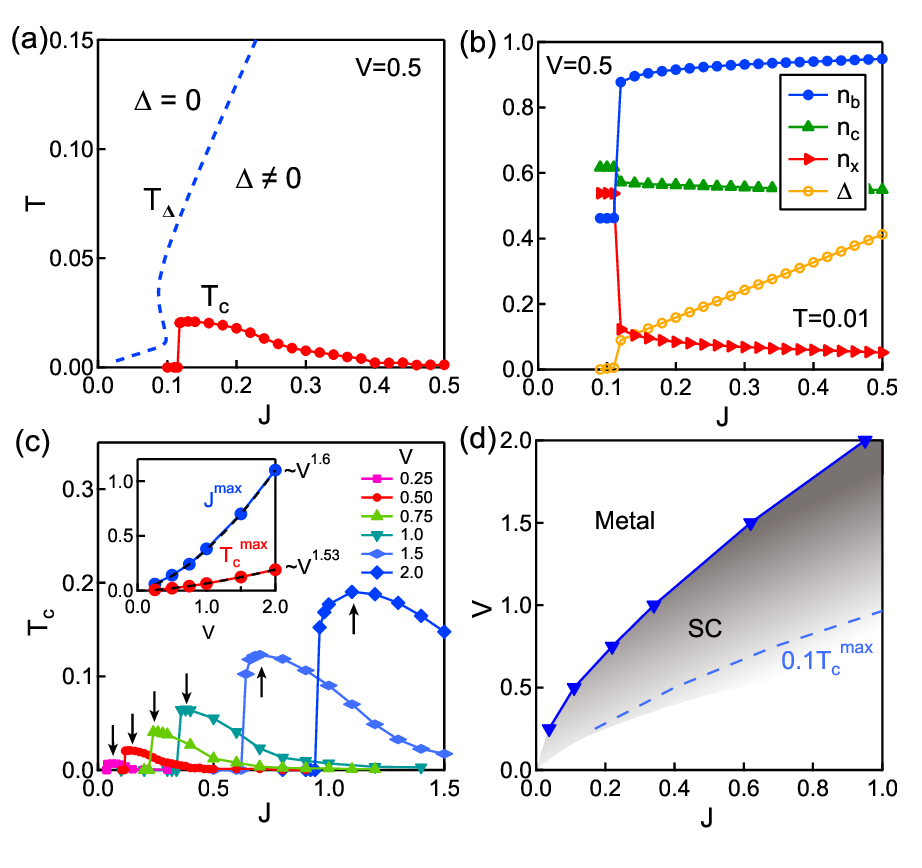}
	\end{center}
	\caption{(a) The finite temperature phase diagram of the $t$-$V$-$J$ model at  $V=0.5$. The dashed line is the crossover temperature $T_\Delta$ marking the formation of the interlayer $d_{z^2}$ spinon valence bonds, and the red line is the superconducting transition temperature $T_c$.  (b) Variation of $\Delta$ and the occupation number of different particles ($n_b$, $n_\chi$, $n_c$) as functions of $J$ for $V=0.5$ and $T=0.01$ calculated from the normal state solution. (c) The superconducting transition temperature $T_c$ plotted as a function of $J$ for different values of $V$. The arrows mark the position $J^\text{max}$ at the maximum transition temperature $T_c^\text{max}$. The inset shows $T_c^\text{max}$ and $J^\text{max}$ as functions of $V$, where the dashed lines are power-law fittings. (d)The ground state phase diagram on the $V$-$J$ plane. The blue solid line is the phase boundary between the metal and superconducting phase (shaded region), and the dashed line is determined by $T_c=0.1T_c^\text{max}$ for each $V$.}
	\label{fig:PD}
\end{figure}

\section{Results}
\textit{Superconducting phase diagram.---}Figure \ref{fig:PD}(a) shows the finite temperature $T$-$J$ phase diagram for a typical hybridization $V=0.5$. As the temperature decreases, the spinons first form local interlayer valence bonds marked by a nonzero $\Delta$ below the crossover temperature $T_\Delta$. Superconductivity then develops at much lower temperatures $T_c$. For large $J$, as shown in Fig. \ref{fig:PD}a, $T_c$ decreases as $T_\Delta$ increases. The opposite trend suggests a dual role of the superexchange $J$. While $J$ is necessary for the superconductivity, the Cooper instability of $d_{x^2-y^2}$-electron pairs may be suppressed if the local pairing is too strong. This is consistent with the two-component theory \cite{YFYang2023,QQin2023}, where the phase coherence may be difficult to achieve if $J$ is too large for a fixed $V$. Below $J_c\approx 0.12$, $T_c$ jumps to zero following a quantum phase transition towards a non-superconducting metallic phase. The first-order nature of the transition here could be an artifact of the Schwinger boson approach, which can be overcome and turned into a continuous one by introducing a small biquadratic Heisenberg term \cite{Wang2021,Wang2022a,Wang2022b,FCZhang2002}. In any case, the ground state properties on both sides of the transition are not affected, which suggests a minimal interlayer coupling required for the superconducting order. This differs from previous static auxiliary field Monte-Carlo simulations of the $t$-$V$-$J$ model, where the superconductivity develops even at small $J$ possibly due to the static approximation \cite{QQin2023}.

To clarify the nature of the quantum phase transition at $J_c$, we calculate the low-temperature ($T=0.01$) occupation numbers of different particles and $\Delta$ from the normal state solution. The results are shown in Fig. \ref{fig:PD}(b) as functions of $J$. As $J$ increases across $J_c$, the holon occupation $n_\chi$ jumps from above $0.5$ to around $0.1$ and then continues to decrease. The $d_{x^2-y^2}$-electron occupation $n_c$ follows the same trend but the jump at $J_c$ is much smaller, while the spinon occupation $n_b$ shows an opposite trend due to the constraint $n_\chi+n_b=1$. Since $n_b$ is proportional to the spin size of $d_{z^2}$ electrons, $S=n_b/2$, the above results suggest a dramatic reduction of the $d_{z^2}$ moment associated with its delocalization at the quantum phase transition. As shown in Fig. \ref{fig:PD}(b), the reduced $d_{z^2}$ moment naturally leads to a strongly suppressed $\Delta$, which in turn detroys the superconductivity. These are quite similar to the Kondo-Heisenberg model, where the hybridization competes with the Heisenberg interaction, leading to a quantum phase transition between a Kondo-screened phase with an enlarged Fermi volume and a magnetically ordered phase \cite{Wang2020,Wang2021,Wang2022a,Wang2022b}. This is certainly expected considering the similarity between the two models.

\begin{figure}[t]
	\begin{center}
		\includegraphics[width=8cm]{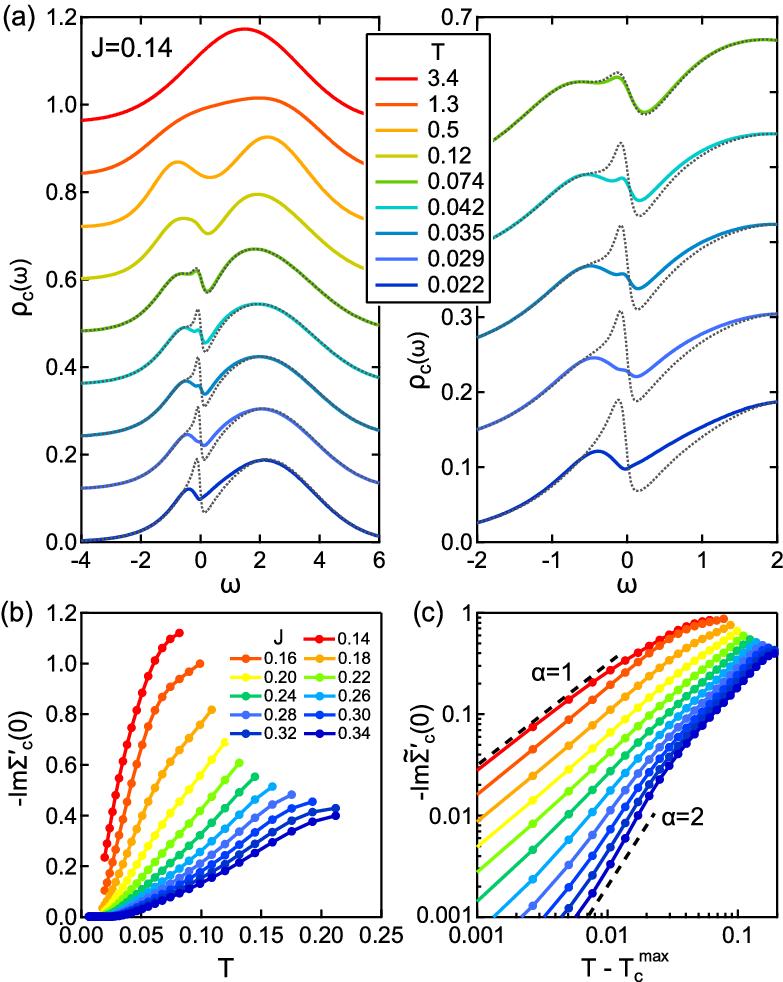}
	\end{center}
	\caption{(a) Temperature evolution of the DOS of $d_{x^2-y^2}$ electrons at $V=0.5$ and $J=0.14$ (left) and the enlarged view around the Fermi energy for $T<T_\Delta$ (right). The dotted curves are the DOS calculated from the $\Delta=0$ solution below $T_\Delta$. For clarity, the curves for different temperatures are shifted by a constant. (b) Temperature dependence of the imaginary part of the $d_{x^2-y^2}$-electron self-energy at zero frequency,  $-\text{Im}\Sigma_c'(0)$, for $V=0.5$ and different values of $J$. (c) Imaginary part of the normalized self-energy, $\tilde{\Sigma}_c'(0)\equiv\Sigma_c'(0,T)-\Sigma_c'(0,T_c^\text{max})$, versus $T-T_c^\text{max}$ for each $J$, where $T_c^\text{max}=0.021$ is the transition temperature at $J=0.14$ and $V=0.5$. The dashed lines correspond to the power-law indices $\alpha=1$ and $\alpha=2$.}
	\label{fig:NS}
\end{figure}

To obtain the phase diagram, we calculate $T_c$ for different values of $V$ and compare the results in Fig. \ref{fig:PD}(c). One can see clearly that $T_c$ scales with $V$. As plotted in the inset, the maximum $T_c^\text{max}$ increases roughly as $V^{1.53}$, and the corresponding optimal $J^\text{max}$ scales roughly as $V^{1.6}$. For $V=0.5$, our result gives $T_c^{\text{max}}\approx 0.021$, corresponding to $117$ K if we take the DFT value of the hopping parameter, $t=0.48$ \cite{DXYao2023}. This is in good consistency with the experimental observation of about 80 K, considering the mean-field treatment of $\Delta$ in our method and the renormalization effect of the hybridization by the Hund's rule coupling \cite{YYCao2024}. The similar scaling also suggests an approximate linear relation in this parameter range, $T_c^\text{max}\approx 0.17 J^\text{max}$, based on our Schwinger boson approach and the Bethe-Salpeter equation, which is also higher than the Monte Carlo results where the superconducting phase fluctuations are better treated \cite{QQin2023}. Figure \ref{fig:PD}(d) gives the ground state phase diagram of the $t$-$V$-$J$ model, where the solid line represents the  transition between the metal and the superconducting phase (shaded region). Since $T_c$ decreases slowly at large $J$, the right boundary of the superconducting phase is hard to determine but marked by the position where $T_c$ is suppressed to $0.1T_c^\text{max}$ for each $V$. Clearly, while the hybridization $V$ and the interlayer superexchange interaction $J$ are both necessary for the superconductivity, we see a subtle competition between them such that the superconductivity is most favored in the middle region of the whole diagram.

\textit{Pseudogap and strange metallicity.---}Above $T_c$, the normal state properties can be studied by calculating the DOS of $d_{x^2-y^2}$ electrons using the maximum-entropy analytic continuation. Figure \ref{fig:NS}(a) shows its evolution from high temperature (red) to low temperature above $T_c=0.021$ (blue) for $V=0.5$ and $J=0.14$.  For comparison, the DOS calculated from the $\Delta=0$ ($J=0$) solution is also shown below the crossover temperature $T_\Delta\approx 0.09$ (dotted curves). For $\Delta=0$, one can see the gradual development of a hybridization gap above the Fermi energy and a sharp quasiparticle peak slightly below the Fermi energy, which is also present in the DFT plus dynamical mean-field theory (DMFT) calculations \cite{YYCao2024}. Direct comparison shows a spectral weight transfer in the $\Delta\neq 0$ solution below $T_\Delta$ from the quasiparticle peak to the hybridization gap, which leads to a ``pseudogap'' feature around the Fermi energy and may be examined in future optical measurement. We attribute this pseudogap to the formation of interlayer spinon pairs, which are responsible for preformed Cooper pairs of $d_{x^2-y^2}$ electrons through the hybridization.

\begin{figure}[t]
	\begin{center}
		\includegraphics[width=8.2cm]{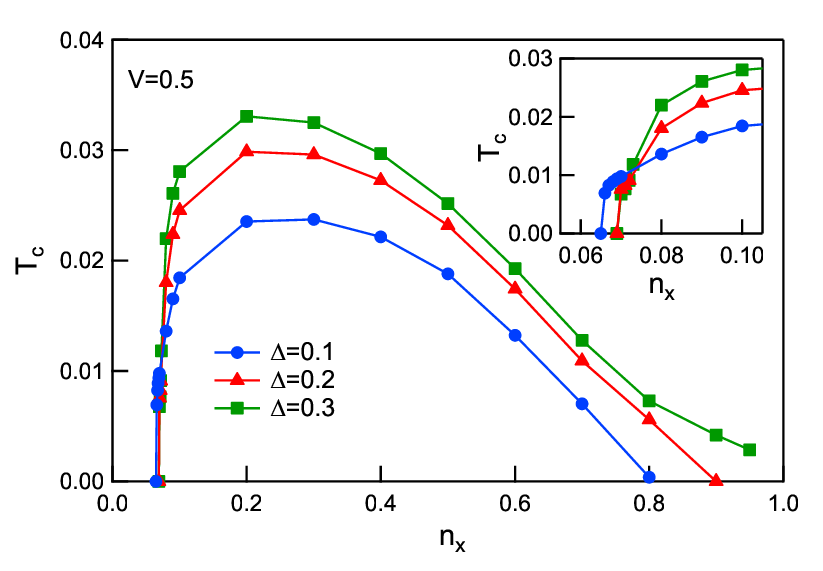}
	\end{center}
	\caption{Evolution of $T_c$ as a function of the holon density $n_\chi$ for $V=0.5$ and different values of $\Delta$, showing a minimal holon density required for the formation of superconductivity. The inset shows an enlarged view around the left boundary of the superconducting phase.}
	\label{fig:nx}
\end{figure}

The strange metal behavior observed experimentally in the normal state of LNO is another important feature associated with strong electronic correlations \cite{HQYuan2024,HHWen2023,LYang2024b}. In our approach, this is caused by the hybridization induced  spinon-holon-electron scattering along the $\mathbf{k}=(0,\pi)$ and $(\pi,0)$ directions, which can be studied from the frequency dependence of $\Sigma_c'(\omega )\equiv \Sigma_c(\mathbf{k},\omega )/\xi_{\bf k}^2$. Figure \ref{fig:NS}(b) shows the temperature dependence of $-\text{Im}\Sigma_c'(\omega=0)$ for $V=0.5$ and different values of $J$. We find a linear-in-$T$ behavior right above $T_c$ at the optimal $J=0.14$, which gradually evolves into a $T^\alpha$ dependence with $1<\alpha<2$ as $J$ increases. For large enough $J$, $-\text{Im}\Sigma_c'(0)$ approaches zero roughly with a $T^2$ dependence at low temperatures, indicating Fermi liquid properties. This can be seen more clearly in Fig. \ref{fig:NS}(c), where the imaginary part of the normalized self-energy $\tilde{\Sigma}_c'(0)\equiv\Sigma_c'(0,T)-\Sigma_c'(0,T_c^\text{max})$ is shown in the double-logarithmic plot as a function of  $T-T_c^\text{max}$, with the maximum transition temperature $T_c^\text{max}=0.021$ at $J=0.14$ being the reference temperature. The evolution from the strange metal to Fermi liquid behaviors can be easily understood from the competition between the interlayer spin interaction and the hybridization. A sufficiently large $J$ suppresses the hybridization effect and causes the Fermi liquid properties of weakly hybridized $d_{x^2-y^2}$ electrons. For the same reason, we find that the pseudogap feature is also gradually suppressed and becomes less obvious as $J$ increases (see Appendix \ref{app}).

\textit{The effect of $d_{z^2}$ holes.---}One of the key differences between the hybridization driven and the Hund coupling driven superconductivity lies in the $d_{z^2}$-hole dependence of $T_c$. The hybridization scenario requires the $d_{z^2}$ charge (holon) fluctuations to transfer the local interlayer spinon pairs to the mobile $d_{x^2-y^2}$-electron Cooper pairs, suggesting a finite holon density may be beneficial for the superconductivity. To study the effect of $d_{z^2}$ holes on the superconducting transition, we introduce the holon chemical potential $\mu_\chi$ to tune its occupation number $n_\chi$. For simplicity, here we fix the spinon pairing amplitude $\Delta$ as an input parameter and focus solely on the effect of $n_\chi$. Figure \ref{fig:nx} shows the $n_\chi$ dependence of $T_c$ for $V=0.5$ and $\Delta=0.1-0.3$. As $n_\chi$ increases, $T_c$ suddenly jumps (weakly first-order) to a finite value at $n_\chi\approx 0.065-0.07$, quickly reaches its maximal value at $n_\chi\approx 0.2$, and then slowly decreases towards zero at large $n_\chi$. The suppression of $T_c$ for sufficiently large $n_\chi$ is due to the small spinon occupation $n_b=1-n_\chi$, which results in a significantly reduced number of spinon pairs that can no longer support the superconductivity. On the other hand, the vanishment of $T_c$ at small $n_\chi< 0.065$ suggests a minimal $d_{z^2}$-hole concentration for the superconducting order, consistent with the experimental observation that the superconductivity in LNO occurs only in the high pressure phase where the $\gamma$ band of $d_{z^2}$ orbital is lifted to the Fermi level and acquires a hole pocket \cite{MWang2023Nature,DXYao2023}. The overall $T_c$ curves obtained here are highly asymmetric, which has also been obtained previously for the two-dimensional Hubbard model near the Mott transition using the plaquette dynamical mean-field theory and ascribed to the remnant of the first-order transition between the pseudogap phase and the correlated metal phase \cite{Fratino2016}. Their shape looks similar to the right-triangle shape reported experimentally in LNO, where it has been argued to be caused by the structure transition \cite{MWang2023Nature,JGCheng2023,HQYuan2024,MWang2024b}. The similarity here raises the question concerning the true driving force of this transition.

\section{Discussions and conclusion}

We have studied the effective bilayer $t$-$V$-$J$ model of LNO using the Schwinger boson (slave fermion) representation of the strongly correlated $d_{z^2}$ electrons. Within this formalism, Cooper pairs of $d_{x^2-y^2}$ electrons can be induced by the interlayer spinon valence bonds through charge (holon) fluctuations of the $d_{z^2}$ electrons due to their hybridization. We develop an efficient method to calculate the Cooper instability and obtain a global phase diagram with a hybridized metal and a superconducting ground state separated by a  quantum phase transition. The normal state above the superconducting transition is found to exhibit a pseudogap feature associated with preformed Cooper pairs and non-Fermi liquid or strange metal behaviors due to the hybridization-induced spinon-holon-electron scattering. We further predict a minimal hole concentration of the $d_{z^2}$ orbital needed to establish the superconducting order. These results are in stark contrast to the Hund scenario, where the superconductivity is rather enhanced as the $d_{z^2}$ electrons are tuned towards the half-filled Mott insulating state \cite{GSu2023b,CJWu2023}. Our interpretation of the role of the $\gamma$ pocket is also different from the spin fluctuation theories based on Fermi surface nesting \cite{FYang2023,QHWang2023}. It may therefore be possible to clarify which scenario is correct by experimentally investigating the influence of $d_{z^2}$ holes on the superconducting transition of pressurized LNO.

Our $t$-$V$-$J$ description of LNO is also different from the Anderson or Kondo lattice description of typical heavy fermion systems. The two have very similar mathematical formalism in the Schwinger boson representation and some physical results are also comparable. However, from the theoretical point of views, there are two major differences. One is associated with the hole doping of the localized orbital. While the $d_{z^2}$ holes are crucial here for the superconductivity of LNO, many heavy fermion superconductors, in particular the Ce-based ones, contain fully localized $f$-orbitals and their superconductivity is primarily considered to be associated with the Kondo resonance state, which is a higher order effect compared to LNO with self-doped holes on $d_{z^2}$ orbitals. Our theory suggests that it may also be interesting to investigate the superconductivity in the mixed valence regime of the Anderson lattice, which may actually be more suitable for the U-based superconductors. The second major distinction involves the origin of the pairing interaction. The interlayer superexchange interaction in the bilayer LNO is much more effective in providing strong pairing strength than the Rudermann-Kittel-Kasuya-Yosida (RKKY)  interaction in heavy fermion systems, since the latter often suffers from intrinsic frustration associated with pairing one spin with many other nearby spins. This is in some sense analogous to the case of La$_4$Ni$_3$O$_{10}$, where each $d_{z^2}$ spin in the inner layer simultaneously couples with two spins in the outer layers and leads to a weaker pairing amplitude and a smaller $T_c$ \cite{QQin2024}. Moreover, the RKKY interaction is typically much smaller than the superexchange interaction. These explain the much lower superconducting transition temperatures in typical heavy fermion superconductors. On the other hand, the hybridization between $d_{x^2-y^2}$ and $d_{z^2}$ orbitals in LNO may also give rise to an effective intralayer RKKY interaction, whose interplay with the onsite Hund's rule coupling may be responsible for the magnetic orders observed in experiments at ambient pressure.
 
\acknowledgements
This work was supported by the National Natural Science Foundation of China (Grants No. 12304174 and No. 12174429), the Strategic Priority Research Program of the Chinese Academy of Sciences (Grant No. XDB33010100), and the National Key Research and Development Program of China (Grant No. 2022YFA1402203).

\begin{figure}[t]
	\begin{center}
		\includegraphics[width=8cm]{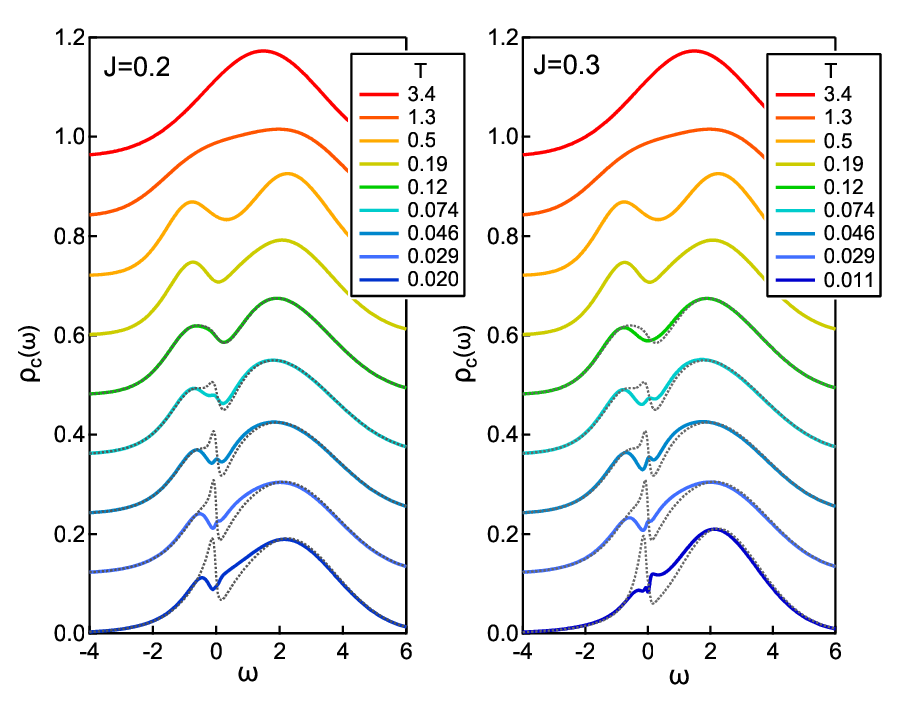}
	\end{center}
	\caption{Temperature evolution of the DOS of $d_{x^2-y^2}$ electrons for $V=0.5$ and $J=0.2$ (left), $0.3$ (right). The dashed curves are the DOS calculated from the $\Delta=0$ solution below $T_\Delta$. The curves for different temperatures are shifted by a constant for clarity. }
	\label{fig5}
\end{figure}

\appendix

\section{Pseudogap at large $J$}\label{app}
For completeness, we show in Fig. \ref{fig5} the temperature evolution of the $d_{x^2-y^2}$-electron DOS for $V=0.5$ and $J=0.2$, $0.3$. One can still see the spectral weight transfer from the quasiparticle peak to the hybridization gap and the development of the pseudogap around the Fermi energy. However, for $J=0.3$, additional shoulder features arise around the Fermi energy at low temperature, making the pseudogap feature less obvious. The overall shape of DOS also becomes closer to the bare one at $V=0$ (or the one at very high temperature), indicating the effect of hybridization is strongly suppressed at such a large value of $J$.


\begin{thebibliography}{99}
	
	\bibitem{MWang2023Nature}H. Sun, M. Huo, X. Hu, J. Li, Z. Liu, Y. Han, L. Tang, Z. Mao, P. Yang, B. Wang, J. Cheng, D.-X. Yao, G.-M. Zhang, and M. Wang, Signatures of superconductivity near 80 K in a nickelate under high pressure, Nature \textbf{621}, 493 (2023).
	
	\bibitem{JGCheng2023}J. Hou, P. T. Yang, Z. Y. Liu, J. Y. Li, P. F. Shan, L. Ma, G. Wang, N. N. Wang, H. Z. Guo, J. P. Sun, Y. Uwatoko, M. Wang, G.-M. Zhang, B. S. Wang, and J.-G. Cheng, Emergence of high-temperature superconducting phase in the pressurized La$_3$Ni$_2$O$_7$ crystals,
	Chin. Phys. Lett. \textbf{40}, 117302 (2023).
	
	\bibitem{HQYuan2024}Y. Zhang, D. Su, Y. Huang, Z. Shan, H. Sun, M. Huo, K. Ye, J. Zhang, Z. Yang, Y. Xu, Y. Su, R. Li, M. Smidman, M. Wang, L. Jiao, and H. Yuan, High-temperature superconductivity with zero-resistance and strange-metal behaviour in La$_{3}$Ni$_{2}$O$_{7-\delta}$, Nat. Phys. {\bf 20}, 1269 (2024).
	
	\bibitem{HHWen2023}Z. Liu, M. Huo, J. Li, Q. Li, Y. Liu, Y. Dai, X. Zhou, J. Hao, Y. Lu, M. Wang, and H.-H. Wen, Electronic correlations and partial gap in the bilayer nickelate La$_{3}$Ni$_{2}$O$_{7}$, Nat. Commun. {\bf 15}, 7570 (2024).
	
	\bibitem{XJZhou2024}J. Yang, H. Sun, X. Hu, Y. Xie, T. Miao, H. Luo, H. Chen, B. Liang, W. Zhu, G. Qu, C.-Q. Chen, M. Huo, Y. Huang, S. Zhang, F. Zhang, F. Yang, Z. Wang, Q. Peng, H. Mao, G. Liu, Z. Xu, T. Qian, D.-X. Yao, M. Wang, L. Zhao, and X. J. Zhou, Orbital-dependent electron correlation in double-layer nickelate La$_3$Ni$_2$O$_7$, Nat. Commun. {\bf 15}, 4373 (2024).
	
	\bibitem{LShu2024}K. Chen, X. Liu, J. Jiao, M. Zou, C. Jiang, X. Li, Y. Luo, Q. Wu, N. Zhang, Y. Guo, and L. Shu, Evidence of spin density waves in La$_3$Ni$_2$O$_{7-\delta}$, Phys. Rev. Lett. {\bf 132}, 256503 (2024).
	
	\bibitem{ZChen2024}Z. Dong, M. Huo, J. Li, J. Li, P. Li, H. Sun, Y. Lu, M. Wang, Y. Wang, and Z. Chen, Visualization of oxygen vacancies and self-doped ligand holes in La$_3$Ni$_2$O$_{7-\delta}$, Nature {\bf 630}, 847 (2024).
	
	\bibitem{XChen2024}Z. Dan, Y. Zhou, M. Huo, Y. Wang, L. Nie, M. Wang, T. Wu, and X. Chen, Spin-density-wave transition in double-layer nickelate La$_3$Ni$_2$O$_{7}$, arXiv:2402.03952.
	
	\bibitem{ZXShen2024}S. N. Abadi, K. Xu, E. G. Lomeli, P. Puphal, M. Isobe,  Y. Zhong, A. V Fedorov, S. Mo, M. Hashimoto, D. Lu, B. Moritz, B. Keimer, T. P. Devereaux, M. Hepting, and Zhi-Xun Shen, Electronic structure of the alternating monolayer-trilayer phase of La$_3$Ni$_2$O$_7$, arXiv:2402.07143.
	
	
	
	\bibitem{JGCheng2024PRX}G. Wang, N. N. Wang, X. L. Shen, J. Hou, L. Ma, L. F. Shi, Z. A. Ren, Y. D. Gu, H. M. Ma, P. T. Yang, Z. Y. Liu, H. Z. Guo, J. P. Sun, G. M. Zhang, S. Calder, J.-Q. Yan, B. S. Wang, Y. Uwatoko, and J.-G. Cheng, Pressure-Induced superconductivity in polycrystalline ${\mathrm{La}}_{3}{\mathrm{Ni}}_{2}{\mathrm{O}}_{7\ensuremath{-}\ensuremath{\delta}}$, Phys. Rev. X \textbf{14}, 11040 (2024).
	

	
	\bibitem{DLFeng2024}X. Chen, J. Choi, Z. Jiang, J. Mei, K. Jiang, J. Li, S. Agrestini, M. Garcia-Fernandez, X. Huang, H. Sun, D. Shen, M. Wang, J. Hu, Y. Lu, K.-J. Zhou, and D. Feng, Electronic and magnetic excitations in La$_3$Ni$_2$O$_7$, arXiv:2401.12657.
	
	\bibitem{MWang2024}T. Xie, M. Huo, X. Ni, F. Shen, X. Huang, H. Sun, H. C. Walker, D. Adroja, D. Yu, B. Shen, L. He, K. Cao, and M. Wang, Neutron scattering studies on the high-$T_c$ superconductor La$_3$Ni$_2$O$_{7-\delta}$ at ambient pressure, arXiv:2401.12635.
	
	\bibitem{MWang2024b} J. Li, P. Ma, H. Zhang, X. Huang, C. Huang, M. Huo, D. Hu, Z. Dong, C. He, J. Liao, X. Chen, T. Xie, H. Sun, and M. Wang, Pressure-driven right-triangle shape superconductivity in bilayer nickelate La$_3$Ni$_2$O$_7$, arXiv:2404.11369.
	
	\bibitem{HZLiu2024} L. Wang, Y. Li, S.-Y. Xie, F. Liu, H. Sun, C. Huang, Y. Gao, T. Nakagawa, B. Fu, B. Dong, Z. Cao, R. Yu, S. I. Kawaguchi, H. Kadobayashi, M. Wang, C. Jin, H.-K. Mao, and H. Liu,  Structure responsible for the superconducting state in La$_3$Ni$_2$O$_7$ at high pressure and low temperature conditions, J. Am. Chem. Soc. {\bf 146}, 7506 (2024).
	
	\bibitem{LYang2024} Y. Li, X. Du, Y. Cao, C. Pei, M. Zhang, W. Zhao, K. Zhai, R. Xu, Z. Liu, Z. Li, J. Zhao, G. Li, Y. Qi, H. Guo, Y. Chen, L. Yang, Electronic correlation and pseudogap-like behavior of high-temperature superconductor La$_3$Ni$_2$O$_7$, Chin. Phys. Lett. {\bf 41}, 087402 (2024).
	
	\bibitem{LYang2024b} Y. D. Li, Y. T. Cao, L. Y. Liu, P. Peng, H. Lin, C. Y. Pei, M. X. Zhang, H. Wu, X. Du, W. X. Zhao, K. Y. Zhai, J. K. Zhao, M.-L. Lin, P. H. Tan, Y. P. Qi, G. Li, H. J. Guo, L. Yang, and L. X. Yang, Ultrafast dynamics of bilayer and trilayer nickelate superconductors, arXiv:2403.05012.
	
	\bibitem{TXiang2024review} M. Wang, H.-H. Wen, T. Wu, D.-X. Yao, and T. Xiang, Normal and superconducting properties of La$_3$Ni$_2$O$_7$, arXiv:2406.04837. 
	
	\bibitem{Yang2024arxiv} Y.-F. Yang, Possible Fano effect and suppression of Andreev reflection in La$_3$Ni$_2$O$_7$, arXiv:2408.14294.
	
	\bibitem{DXYao2023}Z. Luo, X. Hu, M. Wang, W. W\'u, and D.-X. Yao, Bilayer two-orbital model of La$_3$Ni$_2$O$_7$ under pressure, Phys. Rev. Lett. \textbf{131}, 126001 (2023).

    \bibitem{Dagotto2023} Y. Zhang, L.-F. Lin, A. Moreo, and E. Dagotto, Electronic structure, dimer physics, orbital-selective behavior, and magnetic tendencies in the bilayer nickelate superconductor La$_3$Ni$_2$O$_7$ under pressure, Phys. Rev. B {\bf 108}, L180510 (2023).
	
	\bibitem{Werner2023}V. Christiansson, F. Petocchi, and P. Werner, Correlated electronic structure of La$_3$Ni$_2$O$_7$ under pressure, Phys. Rev. Lett. \textbf{131}, 206501 (2023).
	
	\bibitem{Leonov2023}D. A. Shilenko and I. V. Leonov, Correlated electronic structure, orbital-selective behavior, and magnetic correlations in double-layer La$_3$Ni$_2$O$_7$ under pressure, Phys. Rev. B \textbf{108}, 125105 (2023).
	
	\bibitem{Eremin2023}F. Lechermann, J. Gondolf, S. B\"otzel, and I. M. Eremin, Electronic correlations and superconducting instability in La$_3$Ni$_2$O$_7$ under high pressure, Phys. Rev. B \textbf{108}, L201121 (2023).
	
	\bibitem{YYCao2024}Y. Cao and Y.-F. Yang, Flat bands promoted by Hund's rule coupling in the candidate double-layer high-temperature superconductor La$_3$Ni$_2$O$_7$, Phys. Rev. B \textbf{109}, L081105 (2024).
	
	\bibitem{Eremin2024} F. Lechermann, S. B\"otzel, and I. M. Eremin, Electronic instability, layer selectivity, and Fermi arcs in La$_3$Ni$_2$O$_7$, Phys. Rev. Mater. {\bf 8}, 074802 (2024).
	

	
	
	\bibitem{YFYang2023}Y.-F. Yang, G.-M. Zhang, and F.-C. Zhang, Interlayer valence bonds and two-component theory for high-$T_{c}$ superconductivity of La$_{3}$Ni$_{2}$O$_{7}$ under pressure, Phys. Rev. B \textbf{108}, L201108 (2023).
	
	\bibitem{QQin2023}Q. Qin and Y.-F. Yang, High-$T_c$ superconductivity by mobilizing local spin singlets and possible route to higher $T_c$ in pressurized La$_3$Ni$_2$O$_7$, Phys. Rev. B \textbf{108}, L140504 (2023).
	
	\bibitem{GMZhang2023}Y. Shen, M. Qin, and G.-M. Zhang, Effective bi-layer model hamiltonian and density-matrix renormalization group study for the high-$T_c$ superconductivity in La$_3$Ni$_2$O$_7$ under high pressure, Chin. Phys. Lett. \textbf{40}, 127401 (2023).
	
	\bibitem{WWu2023}Y.-Y. Zheng and W. W\'u, Superconductivity in the bilayer two-orbital Hubbard model, arXiv:2312.03605.
	
	\bibitem{DXYao2023tJ}Z. Luo, B. Lv, M. Wang, W. W\'u, and D.-X. Yao, High $T_c$ superconductivity in La$_3$Ni$_2$O$_7$ based on the bilayer two-orbital $t$-$J$ model, npj Quantum Mater. {\bf 9}, 61 (2024).
	
	
	\bibitem{TZhou2023} J. Huang, Z. D. Wang, and T. Zhou, Impurity and vortex states in the bilayer high-temperature superconductor La$_3$Ni$_2$O$_7$, Phys. Rev. B {\bf 108}, 174501 (2023).
	
	\bibitem{CJWu2023}C. Lu, Z. Pan, F. Yang, and C. Wu, Interplay of two E$_g$ orbitals in superconducting La$_3$Ni$_2$O$_7$ under pressure, Phys. Rev. B {\bf 110}, 094509 (2024).
	
	\bibitem{JPHu2023}Y. Gu, C. Le, Z. Yang, X. Wu, and J. Hu, Effective model and pairing tendency in bilayer Ni-based superconductor La$_3$Ni$_2$O$_7$, arXiv:2306.07275.
	
	\bibitem{FYang2023}Y.-B. Liu, J.-W. Mei, F. Ye, W.-Q. Chen, and F. Yang, $s^\pm$-Wave pairing and the destructive role of apical-oxygen deficiencies in La$_3$Ni$_2$O$_7$ under pressure, Phys. Rev. Lett. \textbf{131}, 236002 (2023).
	
	\bibitem{QHWang2023}Q.-G. Yang, D. Wang, and Q.-H. Wang, Possible $s_{\pm}$-wave superconductivity in La$_3$Ni$_2$O$_7$, Phys. Rev. B \textbf{108}, L140505 (2023).
	
	\bibitem{GSu2023}X.-Z. Qu, D.-W. Qu, J. Chen, C. Wu, F. Yang, W. Li, and G. Su, Bilayer $t$-$J$-$J_\perp$ model and magnetically mediated pairing in the pressurized nickelate La$_3$Ni$_2$O$_7$, Phys. Rev. Lett. \textbf{132}, 036502 (2024).
	
	\bibitem{CJWu2024}C. Lu, Z. Pan, F. Yang, and C. Wu, Interlayer coupling driven high-temperature superconductivity in La$_3$Ni$_2$O$_7$ under pressure, Phys. Rev. Lett. \textbf{132}, 146002 (2024).
	
	\bibitem{YHZhang2023}H. Oh and Y.-H. Zhang, Type-II $t$-$J$ model and shared superexchange coupling from Hund’s rule in superconducting La$_3$Ni$_2$O$_7$, Phys. Rev. B \textbf{108}, 174511 (2023).
	
	\bibitem{Bohrdt2023} H. Schl\"omer, U. Schollw\"ock, F. Grusdt, and A. Bohrdt, Superconductivity in the pressurized nickelate La$_3$Ni$_2$O$_7$ in the vicinity of a BEC-BCS crossover, arXiv:2311.03349.
	

	
	\bibitem{GSu2023b}X.-Z. Qu, D.-W. Qu, W. Li, and G. Su, Roles of Hund’s rule and hybridization in the two-orbital model for high-T$_c$ superconductivity in the bilayer nickelate, arXiv:2311.12769.

	
	\bibitem{YZYou2023} D.-C. Lu, M. Li, Z.-Y. Zeng, W. Hou, J. Wang, F. Yang, and Y.-Z. You, Superconductivity from doping symmetric mass generation insulators: application to La$_3$Ni$_2$O$_7$ under pressure, arXiv:2308.11195.
	
	\bibitem{ZYWeng2023} J.-X. Zhang, H.-K. Zhang, Y.-Z. You, and Z.-Y. Weng, Strong pairing originated from an emergent $\mathbb{Z}_2$ Berry phase in La$_3$Ni$_2$O$_7$, Phys. Rev. Lett. {\bf 133}, 126501 (2024).
	

	
	\bibitem{ZYLu2024}Y.-H. Tian, Y. Chen, J.-M. Wang, R.-Q. He, and Z.-Y.Lu, Correlation effects and concomitant two-orbital s$_{\pm}$-wave superconductivity in La$_3$Ni$_2$O$_7$ under high pressure, Phys. Rev. B \textbf{109}, 165154 (2024).
	
	\bibitem{FWang2024}J.-R. Xue and F. Wang, Magnetism and superconductivity in the $t$-$J$ model of La$_3$Ni$_2$O$_7$ under multiband Gutzwiller approximation, Chin. Phys. Lett. {\bf 41}, 057403 (2024).
	
	\bibitem{Kuroki2024}H. Sakakibara, N. Kitamine, M. Ochi, and K. Kuroki, Possible high $T_c$ superconductivity in La$_3$Ni$_2$O$_7$ under high pressure through manifestation of a nearly-half-filled bilayer Hubbard model, Phys. Rev. Lett. \textbf{132}, 106002 (2024).
	
	\bibitem{Kuroki2024b}T. Kaneko, H. Sakakibara, M. Ochi, and K. Kuroki, Pair correlations in the two-orbital Hubbard ladder: Implications for superconductivity in the bilayer nickelate ${\mathrm{La}}_{3}{\mathrm{Ni}}_{2}{\mathrm{O}}_{7}$, Phys. Rev. B \textbf{109}, 045154 (2024).
	

	
	\bibitem{WLi2024}J. Chen, F. Yang, and W. Li, Orbital-selective superconductivity in the pressurized bilayer nickelate La$_3$Ni$_2$O$_7$: An infinite projected entangled-pair state study, Phys. Rev. B {\bf 110}, L041111 (2024).
	
	\bibitem{TXiang2023}Z. Fan, J.-F. Zhang, B. Zhan, D. Lv, X.-Y. Jiang, B. Normand, and T. Xiang, Superconductivity in nickelate and cuprate superconductors with strong bilayer coupling, Phys. Rev. B {\bf 110}, 024514 (2024).
	
	\bibitem{KJiang2024} K. Jiang, Z. Wang, and F. Zhang, High temperature superconductivity in La$_3$Ni$_2$O$_7$, Chin. Phys. Lett. {\bf 41}, 017402 (2024).
	
	\bibitem{WKu2024} R. Jiang, J. Hou, Z. Fan, Z.-J. Lang, and W. Ku, Pressure driven fractionalization of ionic spins results in cupratelike high-T$_c$ superconductivity in La$_3$Ni$_2$O$_7$, Phys. Rev. Lett. {\bf 132}, 126503 (2024).
	
	\bibitem{JZhao2024} Y. Zhu, E. Zhang, B. Pan, X. Chen, D. Peng, L. Chen, H. Ren, F. Liu, N. Li, Z. Xing, J. Han, J. Wang, D. Jia, H. Wo, Y. Gu, Y. Gu, L. Ji, W. Wang, H. Gou, Y. Shen, T. Ying, X. Chen, W. Yang, C. Zheng, Q. Zeng,  J. Guo, and J. Zhao, Superconductivity in trilayer nickelate La$_4$Ni$_3$O$_{10}$ single crystals, Nature {\bf 631}, 531 (2024).
	
	\bibitem{QQin2024} Q. Qin, J. Wang, and Y.-F. Yang, Frustrated Superconductivity in the trilayer nickelate La$_4$Ni$_3$O$_{10}$, arXiv:2405.04340.
	
	\bibitem{Yang2024PRB} Y.-F. Yang, Decomposition of multilayer superconductivity with interlayer pairing, Phys. Rev. B {\bf 110}, 104507 (2024).
	
    \bibitem{Yoshioka1989} D. Yoshioka, Slave-fermion mean field theory of the Hubbard model, J. Phys. Soc. Jpn. {\bf 58}, 1516 (1989).
 
    \bibitem{Long2022} Z. Long, J. Wang, and Y.-F. Yang, Dynamic charge Kondo effect and a slave fermion approach to the Mott transition, Phys. Rev. B {\bf 106}, 195128 (2022).
 
    \bibitem{Long2023} Z. Long, J. Wang, and Y.-F. Yang, Slave fermion interpretation of the pseudogap in doped Mott insulators, Phys. Rev. B {\bf 108}, 155127 (2023).
 
    \bibitem{Wang2020} J. Wang, Y.-Y. Chang, C.-Y. Mou, S. Kirchner, and C.-H. Chung, Quantum phase transition in a two-dimensional Kondo-Heisenberg model: A dynamical Schwinger-boson large-$N$ approach, Phys. Rev. B {\bf 102}, 115133 (2020).
 
    \bibitem{Wang2021} J. Wang and Y.-F. Yang, Nonlocal Kondo effect and quantum critical phase in heavy-fermion metals, Phys. Rev. B {\bf 104}, 165120 (2021).
 
    \bibitem{Wang2022a} J. Wang and Y.-F. Yang, A unified theory of ferromagnetic quantum phase transitions in heavy fermion metals, Sci. China-Phys. Mech. Astron. {\bf 65}, 257211 (2022).
 
    \bibitem{Wang2022b} J. Wang and Y.-F. Yang, Z$_2$ metallic spin liquid on a frustrated Kondo lattice, Phys. Rev. B {\bf 106}, 115135 (2022).
 
    \bibitem{Simons2010} A. Alexander and B. D. Simons,  \textit{Condensed Matter Field Theory}, (Cambridge University Press, Cambridge, England, 2010).
        
    \bibitem{FCZhang2002} T. N. De Silva, M. Ma, and F.-C. Zhang, Pathology of Schwinger boson mean-field theory for Heisenberg spin models, Phys. Rev. B {\bf 66}, 104417 (2002).
    
    \bibitem{Fratino2016} L. Fratino, P. S\'emon, G. Sordi, and A.-M. S. Tremblay, An organizing principle for two-dimensional strongly correlated superconductivity, Sci. Rep. {\bf 6},  22715 (2016).

	
\end{thebibliography}
\end{document}